\documentclass{article}
\IfFileExists{Style_and_Template_for_Preprints__arXiv__bio_arXiv_/arxiv.sty}{\usepackage{Style_and_Template_for_Preprints__arXiv__bio_arXiv_/arxiv}}{\usepackage{arxiv}}
\usepackage[utf8]{inputenc}
\usepackage[T1]{fontenc}
\usepackage{cite}
\usepackage{amsmath,amssymb,amsfonts,mathtools}
\usepackage{graphicx}
\IfFileExists{Style_and_Template_for_Preprints__arXiv__bio_arXiv_/figures/intro_task_concept.png}{\graphicspath{{Style_and_Template_for_Preprints__arXiv__bio_arXiv_/figures/}}}{\graphicspath{{figures/}}}
\usepackage{booktabs}
\usepackage{array}
\usepackage{algorithm}
\usepackage{algpseudocode}
\usepackage{textcomp}
\usepackage{url}
\usepackage[hidelinks]{hyperref}
\usepackage{microtype}
\newcommand{\TopK}{\operatorname{TopK}}
\newcommand{\softmax}{\operatorname{softmax}}
\newcommand{\MHA}{\operatorname{MHA}}
\newcommand{\CE}{\operatorname{CE}}

\title{Residual-Calibrated Multimodal Spatiotemporal Learning for Future Beam Prediction}

\author{
Md Ahadul Haque Suvo\\
Department of Avionics Engineering\\
Aviation and Aerospace University\\
Dhaka, Bangladesh\\
\texttt{suvo\_23024006@aaub.edu.bd}
\And
Rifat Bin Reza\\
Department of Electrical and Electronic Engineering\\
Brac University\\
Dhaka, Bangladesh\\
\texttt{rifat.bin.reza1@g.bracu.ac.bd}
\AND
Saifur Rahman Sabuj\\
Department of Electrical and Electronic Engineering, Brac University\\
Dhaka, Bangladesh\\
Department of Automotive Engineering, Hanyang University\\
Seoul, South Korea\\
\texttt{s.r.sabuj@ieee.org}
}

\begin{document}

\maketitle

\begin{abstract}
We propose a residual calibrated multimodal spatiotemporal framework for future beam prediction on DeepSense 6G Scenario 23. The framework integrates visual observations, location features, and historical beam labels to predict the beam label of the next dataset record. A pretrained Swin-Tiny encoder extracts features from four consecutive RGB frames. Bidirectional cross-attention integrates visual and two dimensional Unit 1 location representations, while a temporal Transformer incorporates historical beam label embeddings to model dependencies across the observation window. A supplementary offline residual head adjusts the frozen supervised logits using immediate correctness rewards and an auxiliary classification loss. A fixed probability mixture of the supervised and residual outputs is selected using validation accuracy. From 11,387 aligned records, we construct 11,255 temporal windows and partition them by capture run into 7,854 training, 1,905 validation, and 1,496 test samples. The supervised predictor achieves Top-1, Top-2, Top-3, and Top-5 test accuracies of 62.83\%, 74.80\%, 79.75\%, and 81.08\%, respectively. The residual predictor achieves corresponding accuracies of 61.76\%, 75.07\%, 79.41\%, and 81.02\%. Residual adjustment increases Top-2 accuracy by 0.27 percentage points, while the supervised predictor retains higher Top-1, Top-3, and Top-5 accuracies. The validation-selected mixture coincides with the residual predictor. The proposed framework combines cross-modal sensing features with explicit beam-history conditioning and a separately trained residual adjustment stage for one-record-ahead beam prediction. Initial evaluation on held-out capture runs characterizes the candidate-ranking performance of both prediction stages and favors the supervised variant for Top-1 selection. These findings motivate further study of historical conditioning and residual refinement.
\end{abstract}


\section{Introduction}
Reliable beam alignment in millimeter-wave links depends on selecting beams that remain suitable as propagation geometry, blockage, and motion evolve. RGB imagery provides scene context and potential obstruction cues; location features provide a compact spatial description; and recent beam labels expose the sequence of beam states recorded before prediction. Combining these inputs can condition a future-beam estimate on both the observed environment and recent link history. Prior studies have already explored sensing-assisted beam prediction, multimodal fusion, semantic scene representations, and temporal tracking in vehicular and aerial settings \cite{imran2023semantic,tian2023multimodal,morais2023position,ahmad2023vision,fabiani2024v2v,yeo2026multimodal}.

More broadly, UAVs also serve as aerial relays and edge-computing nodes in IoT networks, where latency and energy constraints shape resource allocation \cite{sabuj2026partialfl}.

Existing work also addresses future-step prediction and reward-based beam management. For example, Yeo \emph{et al.} use four aligned image--GPS observations with cross-modal attention and an LSTM to predict the beam at the next time step, while Ghassemi \emph{et al.} combine multimodal beam-group prediction with reinforcement-learning-based selection within each group \cite{yeo2026multimodal,ghassemi2024mmt}. These studies establish that neither multimodal fusion nor future-beam prediction alone defines the present research question.

This study asks whether explicit learned embeddings of historical beam labels can be integrated with visual and location sequences in a temporal predictor, and whether a separately trained residual actor--critic head can adjust the supervised logits. Given four consecutive records ending at index $r$, the predictor estimates the recorded beam label at index $r+1$. The forecast is therefore one dataset record ahead.

For each input history, a pretrained Swin-Tiny encoder extracts one feature vector from each RGB frame, and a separate multilayer perceptron encodes the corresponding two-dimensional Unit 1 location vectors. Bidirectional cross-attention enables the aligned visual and location sequences to exchange information. Learned embeddings of the four recorded beam labels and temporal embeddings are combined with the fused sensing features before a temporal Transformer processes the sequence. A supervised classification head predicts the next-record label. A supplementary offline residual calibration component (ORCC), implemented as an A3C-inspired residual actor--critic head operating in an offline one-step setting, receives the frozen supervised representation and logits and learns an additive logit adjustment using immediate correctness rewards, a learned value baseline, and an auxiliary classification loss. A fixed probability mixture is selected by validation accuracy; its coefficient is global rather than input-adaptive.

A closely related current-beam study by Reza \emph{et al.} uses EfficientNet-B3 and staged fusion of image features with deterministic position-derived synthetic and geometric representations \cite{reza2026efficient}. The present formulation instead predicts the next dataset record, explicitly embeds a four-record beam-label history, and evaluates a separately fitted residual-adjustment stage. Reza \emph{et al.}'s synthetic features are deterministic transformations of position, not an independently measured sensing modality. Because the prediction targets, label spaces, and input histories differ, the studies are not treated as numerical baselines for one another.

The main contributions are as follows:
\begin{itemize}
    \item \textbf{Multimodal temporal predictor:} We implement a next-record architecture that combines Swin-Tiny visual features, bidirectional visual--location cross-attention, and temporal Transformer processing.
    \item \textbf{Explicit beam-history conditioning:} Learned embeddings make recent recorded beam states directly available alongside sensing features, providing a mechanism to model their relationship to changes in the observations. The current experiments do not isolate the incremental effect of this conditioning through ablation.
    \item \textbf{Decoupled residual actor--critic adjustment:} ORCC is trained separately on frozen supervised representations and logits using an A3C-inspired offline one-step reward-and-value update with an auxiliary classification loss. This design permits fitting the residual stage without updating the visual encoder; it is a supplementary refinement component, not an online controller.
    \item \textbf{Run-disjoint evaluation:} Complete capture runs are assigned exclusively to the training, validation, or test partition, yielding 7,854, 1,905, and 1,496 windows, respectively. On the held-out runs, the supervised predictor leads at Top-1, Top-3, and Top-5, ORCC leads Top-2 by 0.27 percentage points, and the validation-selected mixture matches ORCC at the reported precision.
\end{itemize}

\begin{figure}[t]
\centering
\includegraphics[width=\linewidth]{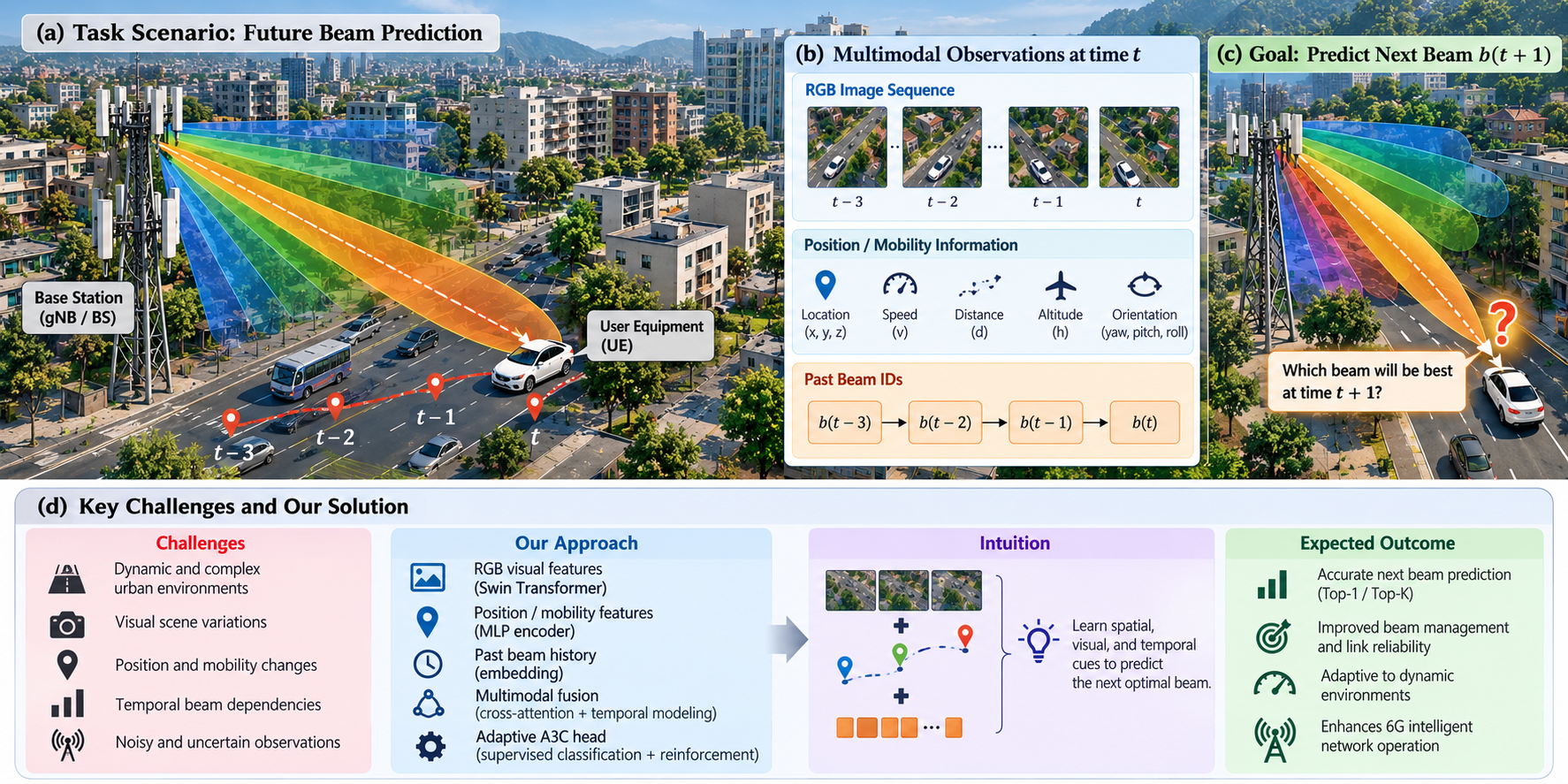}
\caption{Task overview for next-record beam-label prediction. The stationary base station and UAV are schematic context. Four RGB frames, the corresponding two-dimensional Unit 1 location vectors, and four recorded beam labels at indices $r-3$ through $r$ form the input history; the model produces class scores and ranked Top-$K$ candidates for index $r+1$. ORCC is the separately trained, offline one-step A3C-inspired residual actor--critic component.}
\label{fig:intro}
\end{figure}

Figure~\ref{fig:intro} illustrates the prediction task and processing framework. Four consecutive RGB observations, their associated Unit 1 location features, and recorded beam labels form the input history. Cross-modal sensing features and historical beam embeddings are combined through temporal processing to predict the next-record label. A separately trained A3C-inspired residual actor--critic head provides supplementary adjustments to the supervised outputs.

\section{Related Work}
\subsection{Sensing and beam prediction}
DeepSense 6G aligns wireless measurements with sensing streams, supporting research on beam prediction from images, location, LiDAR, radar, and related modalities \cite{alkhateeb2023deepsense,charan2025sensing}. These inputs provide complementary forms of evidence: imagery captures scene appearance and possible obstructions, location describes spatial geometry, and LiDAR or radar can provide range and motion cues. Semantic scene representations offer a higher-level description of visual context, while GPS-aided studies examine position as a practical input whose utility depends on localization quality \cite{imran2023semantic,morais2023position}. Vision-assisted UAV prediction and multimodal learning further demonstrate the breadth of sensing-based beam-index prediction tasks \cite{ahmad2023vision,tian2023multimodal}.

Tian \emph{et al.} process aligned multimodal observations over five sample times with modality-specific feature encoders and a multimodal Transformer, predicting the beam label associated with the terminal sample; their task does not establish a forecast beyond the input sequence \cite{tian2023multimodal}. Ghassemi \emph{et al.} instead use a supervised multimodal Transformer to predict beam groups and then apply Q-learning to choose a beam within the predicted group in a simulated beam-management stage based on DeepSense 6G Scenario 32 \cite{ghassemi2024mmt}. The latter optimizes a reward-based management objective, illustrating how sensor-based classification and sequential beam selection can use related inputs but answer different questions.

\subsection{Multimodal fusion and mobile scenarios}
Fusion and tracking methods have been studied in both vehicular and aerial settings. Camera--GPS fusion has been explored for vehicle-to-vehicle beam tracking, while LiDAR--GPS methods add explicit three-dimensional observations for vehicular tracking \cite{fabiani2024v2v,bian2024lidar}. In UAV communication, Yeo \emph{et al.} combine four-step image and GPS sequences using ResNet-SE visual features, CAformer cross-modal attention, and an LSTM to predict the beam at $t+1$ \cite{yeo2026multimodal}. Chu \emph{et al.} use ground-station imagery and GPS/IMU position and motion features with a ConvNeXt-A visual encoder, efficient channel attention, and late fusion for optimal-beam prediction in a 64-beam setting; a temporal sequence model is not reported \cite{chu2026highprecision}.

The related Scenario 23 study by Reza \emph{et al.} formulates current-beam prediction with a 32-class output and an EfficientNet-B3 visual branch. Its staged image--synthetic--geometry fusion uses deterministic position-derived features, rather than additional independently sensed modalities \cite{reza2026efficient}. In contrast, the present method predicts the next dataset record from four-record histories and makes recorded beam labels explicit inputs to temporal processing. Table~\ref{tab:related-work} compares these tasks and the relevant sensing, fusion, and learning designs; source details that are not specified are marked accordingly.

\begin{table}[t]
\centering
\caption{Comparison of representative sensing-assisted beam prediction and tracking methods.}
\label{tab:related-work}
\small
\setlength{\tabcolsep}{2pt}
\renewcommand{\arraystretch}{1.12}
\begin{tabular}{>{\raggedright\arraybackslash}p{1.65cm} >{\raggedright\arraybackslash}p{2.1cm} >{\raggedright\arraybackslash}p{2.05cm} >{\raggedright\arraybackslash}p{2.75cm} >{\raggedright\arraybackslash}p{3.45cm} >{\raggedright\arraybackslash}p{2.3cm}}
\toprule
Study & Dataset or scenario & Prediction task & Input information & Fusion and temporal model & Learning approach\\
\midrule
Tian et al.~\cite{tian2023multimodal} & DeepSense 6G Challenge; scenarios 31--34 & Predict beam for terminal sample in a five-sample sequence; beyond-input horizon not reported & RGB, LiDAR, and radar sequences; GPS at the first two sample times; beam-label history not reported & Modality-specific CNN encoders and multimodal Transformer over aligned observations & Supervised beam-label prediction\\
Ghassemi et al.~\cite{ghassemi2024mmt} & DeepSense 6G Scenario 32; simulated beam-selection stage & Predict beam group, then select a beam within the group & Image, LiDAR, radar, and GPS; beam-label history not reported & CNN-based multimodal Transformer; detailed temporal history not reported & Supervised group prediction and Q-learning beam selection\\
Yeo et al.~\cite{yeo2026multimodal} & DeepSense 6G UAV setting & Predict beam at $t+1$ from observations $t-3{:}t$ & Four RGB and GPS observations; beam-label history not reported & ResNet-SE, CAformer cross-attention, and LSTM & Supervised beam-index prediction\\
Chu et al.~\cite{chu2026highprecision} & DeepSense 6G UAV setting; 64-beam evaluation & Sample-level optimal-beam prediction; temporal horizon not reported & Ground-station RGB imagery and GPS/IMU position and motion & ConvNeXt-A with ECA and late fusion; temporal sequence model not reported & Supervised beam classification\\
Reza et al.~\cite{reza2026efficient} & DeepSense 6G Scenario 23; 32 classes & Current-beam prediction from a synchronized observation & RGB image and deterministic position-derived features; beam history not used & EfficientNet-B3 with staged image--synthetic--geometry fusion & Supervised classification\\
Proposed framework & DeepSense 6G Scenario 23; 56 observed encoded labels & Predict $b_{r+1}$ from records $r-3{:}r$ & Four RGB frames, two-dimensional Unit 1 locations, and four recorded beam labels & Swin-Tiny, bidirectional visual--location cross-attention, temporal Transformer, and ORCC residual adjustment & Supervised prediction plus offline one-step A3C-inspired actor--critic refinement; fixed validation-selected mixture\\
\bottomrule
\end{tabular}
\end{table}

The comparison clarifies that next-step prediction itself is not unique to this work: Yeo \emph{et al.} also predict a future beam from sensor sequences. Relative to Reza \emph{et al.}'s current-beam classifier, the present formulation changes the target to the next dataset record, explicitly conditions on historical beam labels, and combines visual and location sequences through bidirectional attention before temporal Transformer processing. The separate ORCC stage then learns an additive logit adjustment from frozen representations and supervised logits. These choices are intended to make link history available to the predictor and to decouple residual fitting from visual-encoder training; the current experiments do not isolate gains attributable to either mechanism. Published accuracy scores are not compared as matched baselines because targets, label spaces, inputs, and evaluation protocols differ.

\subsection{Learning objectives and evaluation}
Supervised beam prediction optimizes agreement with recorded labels, whereas reward-based beam management can optimize a system objective such as throughput through sequential decisions. Ghassemi \emph{et al.}, for example, combine supervised beam-group prediction with Q-learning in a simulated beam-selection stage; their policy interacts with a changing state and reward process \cite{ghassemi2024mmt}. ORCC is different: it is an A3C-inspired residual actor--critic head trained offline on fixed contexts of frozen representations and supervised logits. A sampled class receives an immediate correctness reward, a learned value baseline supports the reward-based update, and an auxiliary classification loss is included; the sampled action does not alter subsequent recorded observations. Evaluation in this study reports Top-$K$ accuracy over encoded labels on capture-run-disjoint partitions, not throughput or physical beamforming gain. The observed-history and target-timing assumptions therefore remain central when interpreting results across studies \cite{mnih2016asynchronous,vuckovic2024metrics}.

\section{Problem Formulation}
\subsection{Communication model and assumptions}
For system-level context, consider a generic narrowband link with a uniform linear array (ULA) of $N_t$ antennas and a single-antenna receiver \cite{charan2025sensing}. A physical beam codebook with $N_{\mathcal{F}}$ entries can be represented as $\mathcal{F}=\{f_k\in\mathbb{C}^{N_t}:\lVert f_k\rVert_2=1\}_{k=1}^{N_{\mathcal{F}}}$. The classifier predicts 56 observed encoded labels; their relationship to this illustrative physical codebook is unspecified. A ULA steering vector can be written as
\begin{equation}
 f_k=\frac{1}{\sqrt{N_t}}\left[1,e^{\mathrm{i}\pi\sin\theta_k},\ldots,e^{\mathrm{i}\pi(N_t-1)\sin\theta_k}\right]^{\mathsf T},
 \label{eq:ula-codebook}
\end{equation}
where $\theta_k$ is the steering angle associated with physical codebook entry $k$. This generic model defines communication quantities for context; the reported experiments do not evaluate a physical array, codebook, or channel.

Under a narrowband geometric channel assumption, the channel at record $r$ is
\begin{equation}
 h_r=\sum_{\ell=1}^{L}\alpha_{r,\ell}a(\phi_{r,\ell})\in\mathbb{C}^{N_t},
 \label{eq:channel}
\end{equation}
where $L$ is the number of paths, $\alpha_{r,\ell}\sim\mathcal{CN}(0,\sigma_\ell^2)$ is a path gain, $\phi_{r,\ell}$ is its angle of departure, and $a(\phi)\in\mathbb{C}^{N_t}$ is the array response. For a unit-power symbol $s_r$, unit receiver combining, and noise $n_r\sim\mathcal{CN}(0,N_0)$, the received signal and per-beam SNR are
\begin{equation}
 y_{r,k}=h_r^{\mathsf H} f_k s_r+n_r,\qquad
 \gamma_{r,k}=\frac{|h_r^{\mathsf H}f_k|^2}{N_0}.
 \label{eq:snr}
\end{equation}
The channel-optimal physical codebook entry is $k_r^\star=\arg\max_{k\in\{1,\ldots,N_{\mathcal{F}}\}}\gamma_{r,k}$. In this evaluation, recorded labels remain categorical targets and are not mapped to these physical entries. For a predicted physical entry $\hat k_r$, normalized beamforming gain and achievable rate would be
\begin{equation}
 G_r=\frac{|h_r^{\mathsf H}f_{\hat k_r}|^2}{\max_k|h_r^{\mathsf H}f_k|^2}\in[0,1],~~
 R_r=\log_2(1+\gamma_{r,\hat k_r}).
 \label{eq:system-metrics}
\end{equation}
For a test set of $N$ examples, the corresponding mean communication metrics are
\begin{equation}
 \bar G=\frac{1}{N}\sum_{i=1}^{N}G_i,\qquad
 \bar R=\frac{1}{N}\sum_{i=1}^{N}\log_2(1+\gamma_{i,\hat k_i}).
 \label{eq:mean-system-metrics}
\end{equation}
The reported experiments contain no channel or codebook measurements, so these communication metrics are not computed. Top-$K$ accuracy is reported as ranking over encoded labels, not as a substitute for beamforming gain or achievable rate.

Records are indexed within each capture run. Let $I_r$ denote the RGB image at index $r$, $p_r\in\mathbb{R}^{2}$ the selected \texttt{unit1\_loc} vector, and $b_r\in\mathcal{B}$ the corresponding encoded beam label. An input example ending at index $r$ is
\begin{equation}
  X_r=\bigl(I_{r-3:r},p_{r-3:r},b_{r-3:r}\bigr),
  \qquad y_r=b_{r+1}.
  \label{eq:task}
\end{equation}
The interval notation denotes the four observations at indices $r-3,r-2,r-1,r$. We retain a window only when all five records are consecutive and belong to the same capture run. Thus, the target is one dataset record ahead; timestamps are not included, so elapsed time is not characterized. The location input is the parser-selected two-dimensional \texttt{unit1\_loc} field and is treated as a location feature rather than UAV telemetry. Targets are read from \texttt{unit1\_beam\_index}.

The model outputs a distribution over $C=56$ encoded classes:
\begin{equation}
  f_{\theta}(X_r)=\ell_r\in\mathbb{R}^{C},\qquad
  \hat p_r(c)=\frac{\exp(\ell_{r,c})}{\sum_{k=1}^{C}\exp(\ell_{r,k})}.
  \label{eq:classifier}
\end{equation}
The model outputs a distribution over 56 observed encoded classes. These categorical identifiers are used as labels and are not interpreted as physical beam indices.

\section{Proposed Architecture and Training}
\subsection{Input processing}
Each RGB frame is resized to $224\times224$ pixels and normalized with ImageNet channel statistics. Training-time augmentation comprises color jitter, occasional grayscale conversion, and random erasing; evaluation applies only resizing and normalization. Position-normalization statistics are estimated exclusively from observations in the training windows. For coordinate $d\in\{1,2\}$,
\begin{align}
 \mu_d&=\frac{1}{4|\mathcal{W}_{\mathrm{tr}}|}
   \sum_{i\in\mathcal{W}_{\mathrm{tr}}}\sum_{j=0}^{3}p_{i,j,d},\\
 \sigma_d&=\sqrt{\frac{1}{4|\mathcal{W}_{\mathrm{tr}}|}
   \sum_{i\in\mathcal{W}_{\mathrm{tr}}}\sum_{j=0}^{3}(p_{i,j,d}-\mu_d)^2},\\
 \widetilde p_{i,j,d}&=\frac{p_{i,j,d}-\mu_d}{\max(\sigma_d,10^{-4})}.
 \label{eq:positionnorm}
\end{align}
Because the input windows overlap, an observation may contribute more than once to these statistics, as in the implemented window-level pipeline. The position branch uses only the two-dimensional \texttt{unit1\_loc} field; it does not use \texttt{unit2\_speed}, altitude, distance, or orientation.

\subsection{Visual and position encoding}
For each frame, the pretrained \texttt{swin\_tiny\_\allowbreak patch4\_window7\_224} backbone and global average pooling yield a 768-dimensional representation $s_j$. A shared linear layer followed by LayerNorm and GELU projects this representation to
\begin{equation}
 v_j=\operatorname{GELU}\!\left(\operatorname{LN}(W_v s_j+c_v)\right)
 \in\mathbb{R}^{192}.
 \label{eq:vision}
\end{equation}
The training log records successful loading of pretrained weights. Fine-tuning updates only the final Swin stage and normalization parameters; all earlier backbone parameters remain frozen. Accordingly, attention operates on pooled frame-level vectors rather than spatial patch-token sequences. Global average pooling discards patch-level spatial detail; a patch-token model is a relevant ablation but was not evaluated.

The position branch maps each normalized coordinate vector through a shared $2\rightarrow128\rightarrow192$ multilayer projection:
\begin{align}
 a_j&=\operatorname{GELU}\!\left(\operatorname{LN}(W_{p1}\widetilde p_j+c_{p1})\right),\\
 q_j&=W_{p2}\operatorname{Dropout}_{0.25}(a_j)+c_{p2}\in\mathbb{R}^{192}.
 \label{eq:position}
\end{align}
During training, the complete position sequence is zeroed with probability $0.12$. This sequence-level modality dropout regularizes the position branch; it is not an evaluation of robustness to missing or corrupted location inputs.

\subsection{Bidirectional fusion and temporal encoding}
For query and key/value sequences $A,B\in\mathbb{R}^{n\times192}$, multi-head attention is defined by
\begin{equation}
\begin{aligned}
 S_m&=\frac{(AW_m^Q)(BW_m^K)^\top}{\sqrt{d_h}},\\
 h_m&=\operatorname{softmax}(S_m)BW_m^V,\\
 \MHA(A,B,B)&=\operatorname{Concat}(h_1,\ldots,h_H)W^O.
\end{aligned}
\label{eq:mha}
\end{equation}
Here, $n=4$, $H=4$ denotes the number of attention heads, and $d_h=48$ is the dimension of each head; each attention output is in $\mathbb{R}^{n\times192}$. For each head $m$, $W_m^Q,W_m^K,W_m^V\in\mathbb{R}^{192\times48}$ and $W^O\in\mathbb{R}^{192\times192}$. We apply attention in both directions—visual queries over position features and position queries over visual features—to the four aligned pooled vectors:
\begin{align}
 A^I&=\MHA(V,Q,Q),~~~~  A^P=\MHA(Q,V,V),\\
 r_j&=[v_j+A^I_j\,;\,q_j+A^P_j],\\
 \bar r_j&=\operatorname{LN}(r_j)\in\mathbb{R}^{384},\\
 f_j&=\operatorname{GELU}(W_f\bar r_j+c_f)\in\mathbb{R}^{192},\\
 u_j&=\operatorname{Dropout}_{0.30}(f_j),
 \label{eq:fusion}
\end{align}
Here, $V=(v_0,\ldots,v_3)$ and $Q=(q_0,\ldots,q_3)$ are the visual and position sequences, respectively. The projection has $W_f\in\mathbb{R}^{192\times384}$ and $c_f\in\mathbb{R}^{192}$ and maps their concatenated 384-dimensional representation to 192 dimensions.

Let $e(b_j)\in\mathbb{R}^{192}$ be the learned embedding of the historical beam label, and let $\tau_j$ be the learned temporal embedding. In the reported training configuration, each beam embedding is independently dropped with probability $0.12$:
\begin{equation}
 g_j=u_j+m_j e(b_j)+\tau_j,\qquad m_j\sim\operatorname{Bernoulli}(0.88).
 \label{eq:history}
\end{equation}
At evaluation, all four recorded beam embeddings are retained; the reported metrics use this training-time masking configuration.
A single pre-normalized Transformer encoder layer maps $(g_0,\ldots,g_3)$ to $(o_0,\ldots,o_3)$. It has a width of 192, four attention heads, a 576-dimensional feed-forward block, and dropout of 0.30. The classifier takes the final token representation, $z_r=o_3\in\mathbb{R}^{192}$, as its input. The feed-forward matrices have dimensions $W_1\in\mathbb{R}^{576\times192}$, $c_1\in\mathbb{R}^{576}$, $W_2\in\mathbb{R}^{192\times576}$, and $c_2\in\mathbb{R}^{192}$:
\begin{align}
 x_j&=g_j+\operatorname{Dropout}\bigl(\MHA(\operatorname{LN}(g))_j\bigr),\\
 d_j&=\operatorname{Dropout}\!\left(\operatorname{ReLU}(W_1\operatorname{LN}(x_j)+c_1)\right),\\
 o_j&=x_j+\operatorname{Dropout}(W_2d_j+c_2).
 \label{eq:temporal}
\end{align}
Every token represents a record at or before index $r$; no information from the target record is provided to the encoder.

\begin{figure}[!t]
\centering
\includegraphics[width=\linewidth]{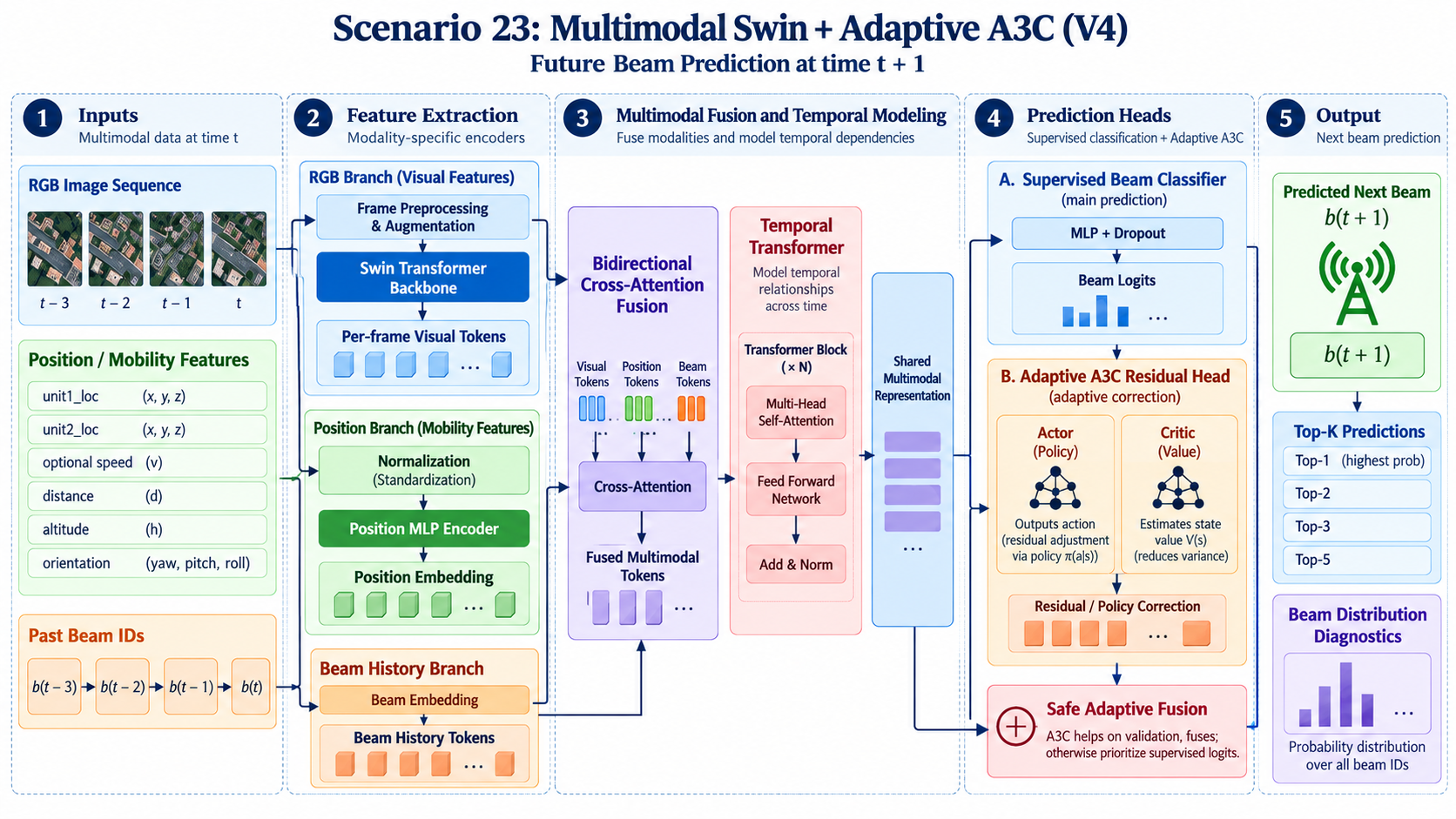}
\caption{Evaluated model architecture. Four frame-level Swin-Tiny features and four two-dimensional Unit 1 location embeddings interact through bidirectional cross-attention. The fused sequence is combined with recorded beam-label and temporal embeddings before the temporal Transformer. The supervised classifier predicts the next-record label; the separately trained ORCC component receives the frozen representation and logits and learns residual logit adjustments. Any probability mixture is a fixed coefficient selected on validation data.}
\label{fig:architecture_concept}
\end{figure}

Figure~\ref{fig:architecture_concept} summarizes the evaluated processing path; the equations and training objectives below specify its implementation.

\subsection{Supervised Learning Objective}
The classification head consists of LayerNorm, dropout with rate $0.45$, a $192\rightarrow96$ linear layer, GELU, dropout with rate $0.30$, and a final $96\rightarrow56$ linear layer. Let $n_c$ be the number of training examples in class $c$, and let $N_{\mathrm{tr}}$ be the total training-set size. We compute class weights from clipped inverse frequencies and normalize them to have unit mean:
\begin{equation}
 \bar w_c=\frac{\operatorname{clip}\!\left(\sqrt{\frac{N_{\mathrm{tr}}}{C\max(n_c,1)}},0.5,3\right)}
 {C^{-1}\sum_{k=1}^{C}\operatorname{clip}\!\left(\sqrt{\frac{N_{\mathrm{tr}}}{C\max(n_k,1)}},0.5,3\right)}.
 \label{eq:weights}
\end{equation}
We use label smoothing with $\epsilon=0.10$, defining the target distribution as $q_{i,c}=(1-\epsilon)\mathbf{1}[y_i=c]+\epsilon/C$. The model minimizes the following weighted cross-entropy loss:
\begin{equation}
 \mathcal{L}_{\mathrm{sup}}=-\frac{1}{\sum_{i=1}^{B}\bar w_{y_i}}\sum_{i=1}^{B}
  \sum_{c=1}^{C}\bar w_c q_{i,c}\log\softmax(\ell_i)_c.
 \label{eq:supervisedloss}
\end{equation}
We optimize with AdamW, using a learning rate of $3\times10^{-6}$ for the unfrozen backbone and $3\times10^{-5}$ for all other parameters, weight decay of $0.05$, and a batch size of 8. Training proceeds for 80 epochs without early stopping; the checkpoint with the highest validation Top-1 accuracy is selected.

\subsection{A3C-inspired residual actor--critic refinement (ORCC)}\label{sec:policy}
The supplementary ORCC component receives the frozen supervised representation and logits, concatenated as $x_i=[z_i;\ell_i]\in\mathbb{R}^{248}$. It is not an independently deployed predictor: its policy adds a learned residual adjustment to the supervised logits. Its shared block is $h_i=\operatorname{LN}(\operatorname{ReLU}(W_hx_i+c_h))\in\mathbb{R}^{128}$, where $W_h\in\mathbb{R}^{128\times248}$ and $c_h\in\mathbb{R}^{128}$. A residual policy distribution and scalar value are then defined by
\begin{align}
 \widetilde\ell_i&=\ell_i+0.25W_\pi h_i+c_\pi, &
 \pi_i&=\softmax(\widetilde\ell_i),\\
 V_i&=w_v^\top h_i+c_v, &
 a_i&\sim\operatorname{Categorical}(\pi_i),\\
 \delta_i&=\mathbf{1}[a_i=y_i]-V_i.
 \label{eq:policy}
\end{align}
Here, $W_\pi\in\mathbb{R}^{56\times128}$, $c_\pi\in\mathbb{R}^{56}$, $w_v\in\mathbb{R}^{128}$, and $c_v\in\mathbb{R}$. The immediate reward is $r_i=\mathbf{1}[a_i=y_i]$, so $\delta_i=r_i-V_i$ is a one-step advantage estimate centered by the learned value baseline. In this paper, residual calibration denotes this learned additive logit adjustment; it does not assert improved probabilistic calibration. The actor--critic objective is
\begin{equation}
\begin{aligned}
 \mathcal{L}_{\mathrm{ORCC}}={}&-\mathbb{E}[\log\pi_i(a_i)\,\operatorname{sg}(\delta_i)]
 +0.25\mathbb{E}[\delta_i^2]\\
 &-0.005\mathbb{E}[H(\pi_i)]
 +0.8\mathbb{E}[\CE(\widetilde\ell_i,y_i)].
\end{aligned}
\label{eq:policyloss}
\end{equation}
where $\operatorname{sg}$ is stop-gradient and $H(\pi_i)=-\sum_c\pi_i(c)\log\pi_i(c)$ is categorical entropy; the auxiliary cross-entropy term is unweighted. Three worker-local actor--critic networks compute updates against shared global ORCC parameters using fixed training examples. Each sampled beam receives the immediate correctness reward above, and the value regression term trains the baseline. Worker computations may overlap, but gradient application and optimizer steps are serialized under a shared lock. The sampled action does not alter later recorded observations or contexts; this is an offline one-step actor--critic update rather than an online interaction loop. We use \emph{A3C-inspired} to describe its worker--global-model design without identifying it as canonical asynchronous A3C \cite{mnih2016asynchronous}. The head is trained for 30 epochs with a shared Adam optimizer at learning rate $2\times10^{-4}$, gradient-norm clipping at 5, and batches of up to 16 fixed training examples. At inference, the residual head produces deterministic class probabilities. The implementation also considers a global mixture coefficient $\alpha\in\{0,0.25,0.5,0.75,1\}$, selected to maximize validation Top-1 accuracy:
\begin{equation}
 p_{\mathrm{mix}}(c\mid X_i)=\alpha\softmax(\ell_i)_c+(1-\alpha)\pi_i(c).
 \label{eq:blend}
\end{equation}
Equation~\eqref{eq:blend} specifies a fixed, global interpolation of class probabilities. At the reported precision, the validation-selected blend and residual-head results have identical test metrics.

\subsection{Training and prediction algorithms}
Algorithms~\ref{alg:supervised}--\ref{alg:inference} summarize the three-stage procedure: supervised classifier fitting and checkpoint selection, residual-head fitting with validation-based selection of the global mixture, and deterministic candidate ranking. The residual stage uses the selected classifier's frozen representations and logits, so it adjusts class scores without updating the base predictor. At inference, the selected components operate with fixed parameters on each held-out history; its target label is used only to calculate final Top-$K$ metrics.

\begin{algorithm}[!ht]
\caption{Run-Disjoint Supervised Training}
\label{alg:supervised}
\begin{algorithmic}[1]
\Require Indexed records with run IDs, RGB frames, \texttt{unit1\_loc}, and beam labels
\Ensure Validation-selected classifier $f_{\theta^*}$
\State Remove unresolved or non-finite rows; order rows within each capture run
\State Form histories and next-record targets as in Eq.~\eqref{eq:task}
\State Reject windows with nonconsecutive indices or run-boundary crossings
\State Assign complete runs chronologically to training, validation, and test partitions
\State Fit position statistics and class weights using Eqs.~\eqref{eq:positionnorm} and \eqref{eq:weights}
\State Initialize pretrained Swin-Tiny multimodal classifier
\For{$e=1$ to $n$}
  \State Optimize the supervised loss in Eq.~\eqref{eq:supervisedloss}
  \State Evaluate validation Top-1 using Eq.~\eqref{eq:topk}
  \If{validation Top-1 exceeds the previous best}
    \State Save the model parameters as $\theta^*$
  \EndIf
\EndFor
\State \Return $f_{\theta^*}$
\end{algorithmic}
\end{algorithm}

\begin{algorithm}[t]
\caption{Offline Residual-Head Fitting and Validation Selection}
\label{alg:policy}
\begin{algorithmic}[1]
\Require Frozen training/validation features and logits; training labels
\Ensure Validation-selected residual head $g_{\psi^*}$ and blend $\alpha^*$
  \State Initialize ORCC head and shared Adam optimizer
\For{$e=1$ to $n$}
  \State Partition the fixed training examples among three workers
  \ForAll{workers in parallel}
    \State Sample batches of at most k examples from the worker's partition
    \State Sample actions and compute centered correctness signals using Eq.~\eqref{eq:policy}
    \State Minimize the ORCC objective in Eq.~\eqref{eq:policyloss}
    \State Backpropagate locally; apply gradients and optimizer step under a shared lock
  \EndFor
  \State Evaluate residual-head Top-1 using Eq.~\eqref{eq:topk}
  \State Save the best validation checkpoint
\EndFor
\State Select $\alpha^*$ by validation Top-1 using Eq.~\eqref{eq:blend}
\State \Return $g_{\psi^*},\alpha^*$
\end{algorithmic}
\end{algorithm}

\begin{algorithm}[t]
\caption{Deterministic Next-Record Prediction and Top-$K$ Evaluation}
\label{alg:inference}
\begin{algorithmic}[1]
\Require History $X_r$ from Eq.~\eqref{eq:task}; selected models; optional target label
\Ensure Ranked beam candidates and optional Top-$K$ score
\State Apply evaluation transforms and position normalization from Eq.~\eqref{eq:positionnorm}
\State Compute representation $z$ and logits $\ell$ using Eqs.~\eqref{eq:temporal} and \eqref{eq:classifier}
\State Compute residual probabilities $\pi$ using Eq.~\eqref{eq:policy}
\State Form the mixture using Eq.~\eqref{eq:blend}
\State Rank classes and score Top-$K$ using Eq.~\eqref{eq:topk}
\If{a held-out label is supplied}
  \State Score whether the target occurs in each returned Top-$K$ set
\EndIf
\end{algorithmic}
\end{algorithm}

\section{Dataset and Experimental Protocol}
The recorded CSV manifest comprises 11,387 rows across 32 sequence or run identifiers. According to the execution log, images and labels were resolved for every row, and the selected position field was valid throughout. Preprocessing yielded 11,255 valid windows. Complete capture runs were assigned chronologically to the training, validation, and test sets, which contain 7,854, 1,905, and 1,496 examples, respectively. Since each run is assigned to a single partition, windows from the same run cannot appear in multiple sets. Table~\ref{tab:split} summarizes the run-disjoint partition sizes and the number of distinct encoded labels in each partition.
\begin{table}[h]
\centering
\caption{Run-disjoint data partition and encoded-label coverage.}
\label{tab:split}
\small
\begin{tabular}{lrr}
\toprule
Partition & Examples & Distinct labels\\
\midrule
Training & 7,854 & 53\\
Validation & 1,905 & 38\\
Test & 1,496 & 31\\
Total vocabulary & 11,255 & 56\\
\bottomrule
\end{tabular}
\end{table}

The label vocabulary was constructed before partitioning and contains 56 encoded classes; the training, validation, and test partitions contain 53, 38, and 31 distinct labels, respectively. Reconstructing the exact split requires the original records and capture-run identifiers. Elapsed time between records is not characterized, and encoded labels are retained as categorical identifiers without a physical beam mapping. Some test labels are absent from training, which limits class-wise interpretation.

The supervised model is trained for 80 epochs, with its checkpoint selected by validation Top-1 accuracy. The supplementary head is trained for 30 epochs and selected by validation argmax accuracy; the same validation set selects the static mixture coefficient. The test partition is used only for final evaluation. Random seeds are specified, but cuDNN benchmarking is enabled and software-environment details are not recorded, so bitwise reproducibility is not assured. Inference latency, computational or energy cost, and variability across random seeds are not measured.

\section{Results and Discussion}
\subsection{Top-\texorpdfstring{$K$}{K} accuracy}
Table~\ref{tab:topk} summarizes the recorded test performance. Given $N$ examples with targets $y_i$ and predicted class probabilities $p_i$, Top-$K$ accuracy is defined as
\begin{equation}
 \operatorname{Acc@K}=\frac{1}{N}\sum_{i=1}^{N}
 \mathbf{1}\!\left[y_i\in\TopK_c p_i(c)\right],~~ K\in\{1,2,3,5\}.
 \label{eq:topk}
\end{equation}
Across the reported test metrics, the supervised classifier leads at Top-1, Top-3, and Top-5, whereas the residual head has a 0.27-percentage-point advantage at Top-2. Relative to the supervised model, its accuracy is lower by 1.07, 0.33, and 0.07 percentage points at Top-1, Top-3, and Top-5, respectively. The validation-selected blend matches the residual-head scores at the reported precision; these results provide no evidence that blending improves performance.

\begin{table}[h]
\centering
\caption{Top-$K$ accuracy on the 1,496-example test set (\%).}
\label{tab:topk}
\small
\begin{tabular}{lrrrr}
\toprule
Method & Top-1 & Top-2 & Top-3 & Top-5\\
\midrule
Supervised Swin & 62.83 & 74.80 & 79.75 & 81.08\\
Offline residual head & 61.76 & 75.07 & 79.41 & 81.02\\
Validation-selected blend & 61.76 & 75.07 & 79.41 & 81.02\\
\bottomrule
\end{tabular}
\end{table}

On the 1,496-example test set, the supervised model places the target in its Top-5 list for 1,213 examples, compared with 1,212 for both ORCC and the blend. Thus, Top-5 coverage differs by only one example despite the residual head's Top-2 advantage; the Top-5 lists still omit 283 supervised-model targets and 284 residual-head targets. These counts describe encoded-label ranking coverage only and do not establish beamforming-gain or rate improvements.

\subsection{Checkpoint selection and validation--test comparison}
Table~\ref{tab:valtest} reports validation and test Top-1 accuracy at the selected checkpoint for each model. The supervised classifier is selected at epoch 67, with validation and test accuracies of 69.03\% and 62.83\%, respectively, yielding a validation-to-test difference of 6.20 percentage points. ORCC is selected at epoch 1; its corresponding accuracies are 69.13\% and 61.76\%, a difference of 7.37 percentage points. These differences are descriptive for this single run-disjoint split and should not be interpreted as uncertainty estimates or as evidence of a statistically significant difference between models.

\begin{table}[t]
\centering
\caption{Validation and test Top-1 accuracy at the selected checkpoints. Accuracy values are percentages; Gap is validation minus test accuracy in percentage points.}
\label{tab:valtest}
\small
\setlength{\tabcolsep}{3pt}
\begin{tabular}{lrrrr}
\toprule
Model & Epoch & Val. Top-1 & Test Top-1 & Gap (pp)\\
\midrule
Supervised Swin & 67 & 69.03 & 62.83 & 6.20\\
ORCC & 1 & 69.13 & 61.76 & 7.37\\
\bottomrule
\end{tabular}
\end{table}

\begin{figure}[h]

\centering
\includegraphics[width=\linewidth]{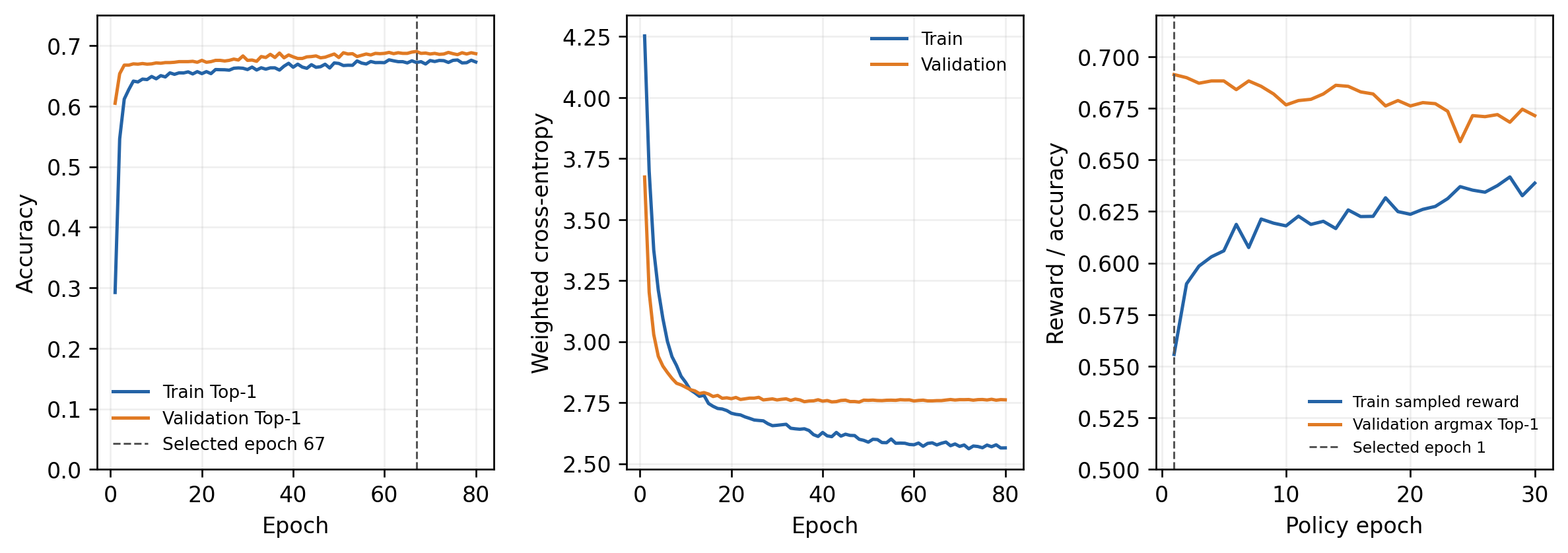}
\caption{Training and validation histories for the supervised classifier and offline residual head. Vertical markers indicate the validation-selected epochs (67 and 1, respectively). Sampled training reward and validation argmax accuracy are distinct measures.}
\label{fig:training}
\end{figure}

\begin{figure}[h]
\centering
\includegraphics[width=0.8\linewidth]{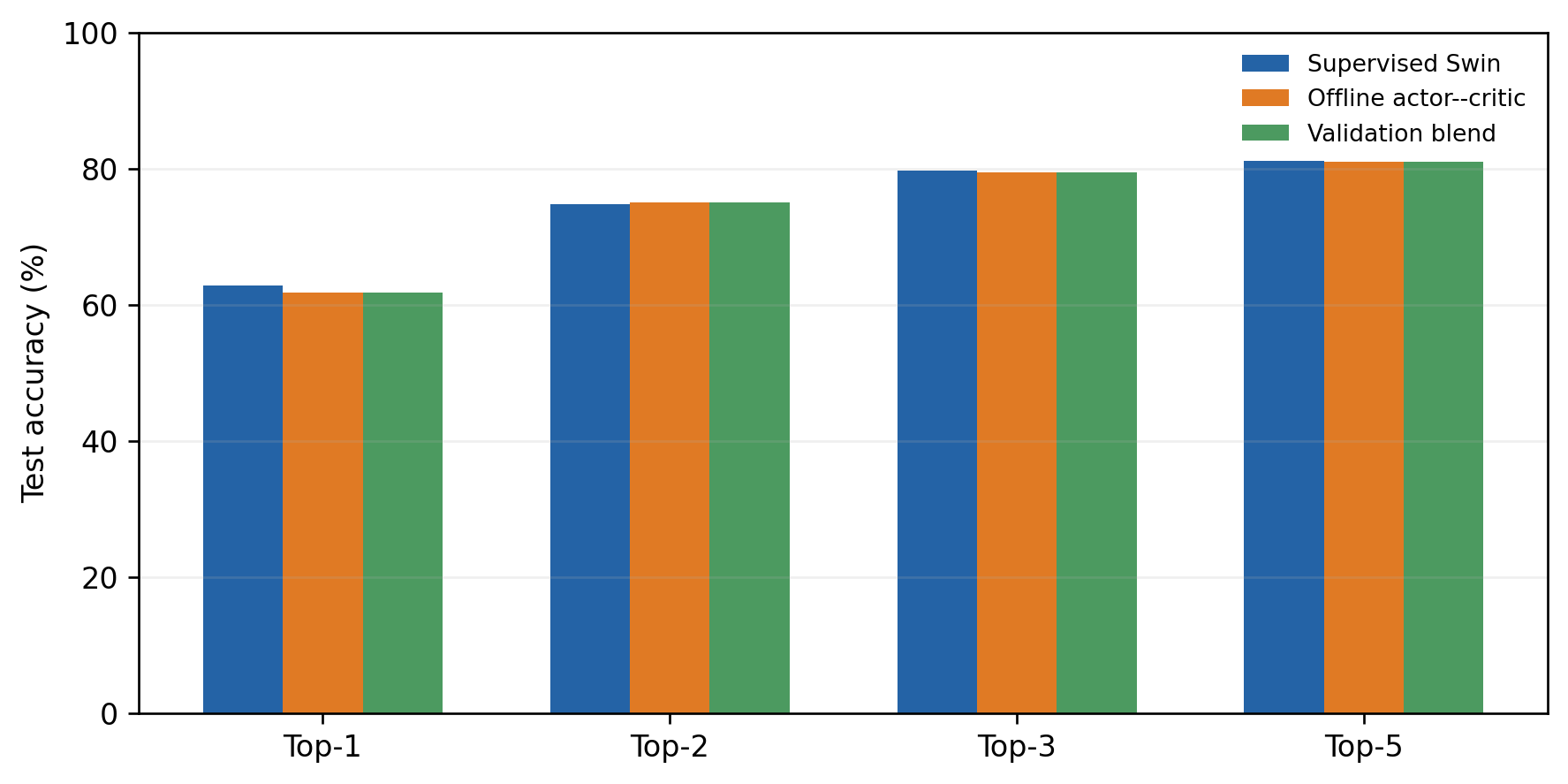}
\caption{Test Top-$K$ accuracy. ORCC and blend scores coincide at the reported precision. ORCC is the offline one-step A3C-inspired residual actor--critic component.}
\label{fig:testtopk}
\end{figure}

Figure~\ref{fig:training} presents the recorded training histories. ORCC's validation argmax accuracy decreases from 69.13\% at epoch 1 to 67.14\% at epoch 30, while its sampled training reward increases from 0.5557 to 0.6387. This divergence is compatible with overfitting but, given the single run and the distinct definitions of reward and validation accuracy, does not establish its cause. Supervised training accuracy is measured under augmentation and is therefore not directly comparable to validation accuracy measured in evaluation mode. Figure~\ref{fig:testtopk} visualizes the test results reported in Table~\ref{tab:topk}.

\subsection{Class-level performance}
The saved classification report yields a macro-F1 of $0.39$ and a weighted-F1 of $0.58$ for the supervised model. Encoded class 31 accounts for 503 test examples (33.6\%) and has an F1 score of $0.93$, whereas class 1 accounts for 241 examples and has an F1 score of $0.00$. This disparity, together with test labels absent from training, cautions against interpreting aggregate accuracy as representative of performance across the full codebook. Encoded identifiers do not correspond to verified physical beam indices.

\subsection{Discussion}
The run-disjoint results characterize the complete implementation on one split; they do not isolate the effects of individual inputs or modules because controlled ablations were not conducted. The literature table is therefore a qualitative comparison of task and method design, not a matched ranking of published accuracies, whose label spaces, inputs, prediction targets, partitions, and metrics differ.

\section{Conclusion}
This study evaluated next-record beam prediction from four RGB frames, four recorded beam labels, and the two-dimensional \texttt{unit1\_loc} field. On one run-disjoint test split, the supervised Swin-Tiny classifier achieved 62.83\% Top-1 accuracy. The separately trained offline one-step A3C-inspired residual actor--critic component exceeded the supervised model only at Top-2, while the fixed validation-selected mixture matched ORCC at the reported precision; residual adjustment therefore did not yield consistent gains. These results characterize ranking over the 56 observed encoded labels rather than communication-level performance. Controlled ablations, repeated runs, and evaluation with a documented physical beam mapping and communication metrics are needed to establish broader benefits.

\bibliographystyle{unsrt}
\IfFileExists{Style_and_Template_for_Preprints__arXiv__bio_arXiv_/references.bib}{\bibliography{Style_and_Template_for_Preprints__arXiv__bio_arXiv_/references}}{\bibliography{references}}
\end{document}